%% file: RJwrapper.tex
\documentclass[a4paper]{report}
\usepackage[utf8]{inputenc}
\usepackage[T1]{fontenc}
\usepackage{RJournal}
\usepackage{amsmath,amssymb,array}
\usepackage{booktabs}
\usepackage{fancyhdr}  % Load the fancyhdr package

\begin{document}

%% do not edit, for illustration only
\sectionhead{Contributed research article}
\volume{XX}
\volnumber{YY}
\year{20ZZ}
\month{AAAA}

%% replace RJtemplate with your article
\begin{article}
  \input{henry}
\end{article}

\bibliography{RJreferences}

\address{Nathan Henry\\
  School of Psychology and Neuroscience\\
  Auckland University of Technology\\
  90 Akoranga Drive, Northcote, Auckland 0627\\
  New Zealand\\
  ORCiD: 0000-0002-4299-1442\\
  \email{nathan.henry@aut.ac.nz}}

\end{document}

%% file: henry.tex
% !TeX root = RJwrapper.tex
\title{CRUD-Capable Mobile Apps with R and shinyMobile: a Case Study in Rapid Prototyping}
\author{by Nathan Henry}

\maketitle

\abstract{
‘Harden’ is a Progressive Web Application (PWA) for Ecological Momentary Assessment (EMA) developed mostly in R, which runs on all platforms with an internet connection, including iOS and Android. It leverages the \pkg{shinyMobile} package for creating a reactive mobile user interface (UI), PostgreSQL for the database backend, and Google Cloud Run for scalable hosting in the cloud, with serverless execution. Using this technology stack, it was possible to rapidly prototype a fully CRUD-capable (Create, Read, Update, Delete) mobile app, with persistent user data across sessions, interactive graphs, and real-time statistical calculation. This framework is compared with current alternative frameworks for creating data science apps; it is argued that the \pkg{shinyMobile} package provides one of the most efficient methods for rapid prototyping and creation of statistical mobile apps that require advanced graphing capabilities. This paper outlines the methodology used to create the Harden application, and discusses the advantages and limitations of the \pkg{shinyMobile} approach to app development. It is hoped that this information will encourage other programmers versed in R to consider developing mobile apps with this framework.
}

\section{Introduction}

\subsection{Mobile app development for statisticians}

Mobile app development is becoming increasingly complex. A statistical programmer faced with the task of creating a mobile app with data science components must compare a growing number of technology stacks to determine the best development method \citep{kasprzak2020six}. If a developer wishes to take full advantage of native capabilities, they will need to learn Java or Kotlin for Android development, and Swift for iOS. If instead, one wishes to create an app that will work on both iOS and Android, they will need to consider a cross-platform Javascript framework such as Ionic or React Native, or perhaps an alternative language such as Flutter. From these two options, a single code base has several advantages, including reduced development and maintenance costs \citep{corral2012potential}. However, each of the aforementioned languages has a steep learning curve, presenting a significant barrier for many in statistical fields, who are often limited in their general programming abilities \citep{chambers2000users}. 

Statistical programmers are often self-taught, or have learned the basics of programming from one or two papers at university. Many lack a fundamental understanding of programming principles (such as the conventions of object-oriented programming) that are generally required for development of complex apps. These factors make the transition from languages such as R to app development languages such as Java or Swift, an intimidating task with a steep learning curve and time barrier. In particular, the prospect of maintaining multiple code bases for different platforms is intimidating, especially for a lone programmer \citep{heitkotter2013cross}. Additionally, while other app development languages may provide access to plotting libraries, none of them are able to match R for its conciseness of statistical code, graphical flexibility, and range of statistical packages. However, it is not possible to install the R runtime on either Android or iOS, making it impossible to incorporate R code into a native mobile application.

\subsection{Progressive Web Apps}

The Progressive Web App (PWA), a set of web standards recently developed by the Google Web Fundamentals group, provides an alternative to native mobile app development frameworks \citep{biorn2017progressive}. A PWA, despite being a web application, can be installed on iOS and Android such that it appears to have native functionality. App icons appear in native format, and the app runs in a fullscreen WebView wrapper. While there are limitations to this format (such as lower performance and lack of native hardware functionality, discussed later in this paper), these are often deemed trivial when considering an app's requirements.

Previously, to build a PWA for mobile, expertise in HTML, CSS and Javascript was required, along with general web development skills. However, the \pkg{shinyMobile} package \citep{granjon2021shinymobile} provides programmers versed in R with the ability to produce fully functional web apps for mobile, with no experience required in any of these languages. This removes a number of barriers for entry into the mobile app market for statistical programmers \citep{elliott2020developing}. As an additional benefit, apps created with \pkg{shinyMobile} also run well on desktop and tablet devices, with the tabbed interface adjusting naturally to different screen sizes. This is largely due to the flexibility of the underlying \pkg{shiny} dashboard framework, which combines with \pkg{shinyMobile} to create a fully reactive app \citep{cheng2021shiny}.

To install a PWA on a mobile phone, one simply needs to open the website in which the app is housed in a supported web browser (such as Google Chrome, Safari, or Mozilla Firefox), then use the browser-specific options menu to add the app icon to their home screen. The app then has the appearance of a native app, with no download required. While the installation process is slightly unconventional when compared to apps installed from Google's Play Store or Apple's App Store, it is straightforward and can be completed within a minute, if instructions are followed correctly. As PWAs become more common, users worldwide are more likely to be familiar with the installation process.

\subsection{The 'Harden' application}

We used the \pkg{shinyMobile} package as our primary framework for developing the ‘Harden’ mobile app, which is designed to help addicts to perform a self-monitored Ecological Momentary Assessment (EMA) to reduce their rate of relapse. This app had several unique requirements:

\begin{enumerate}
    \item Available to mobile users, preferably with near-native appearance, functionality and speed
    \item Advanced graphing with interactive components for self-analysis of variables
    \item Moderately advanced statistical capabilities, such as data wrangling in the backend, and automated calculation of correlations and uncertainties
    \item Intuitive data entry - the user needed to record multiple variables on a daily basis with ease
    \item Straightforward user settings 
    \item Daily notification capabilities
    \item Compliance with relevant security protocols (i.e. HIPAA, GDPR, \textit{et cetera.})
    \item Persistent data across user sessions - users needed the ability to view data they entered into the app on previous days
    \item Authentication, so users could only access their own data.
\end{enumerate}

By using the \pkg{shinyMobile} package to turn this project into a production-ready PWA, all but one of the requirements above were met directly, with notifications not being generated directly by \pkg{shinyMobile}. Workarounds for this are discussed in the \textbf{Limitations} section, below.

\begin{figure}[!ht]
    \begin{center}
        \includegraphics[width=0.7\textwidth]{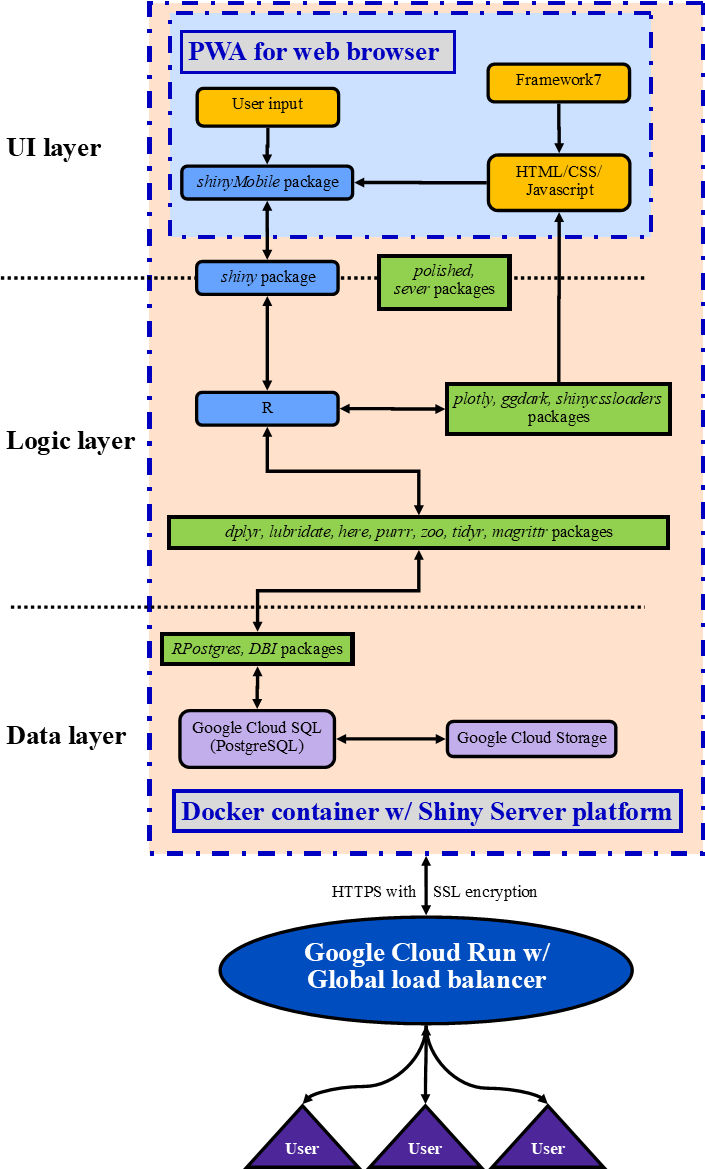}
        \caption{Recommended Harden app architecture. The transition to Google Cloud SQL will be performed in a future version of the app.}
        \label{fig:harden_architecture}
    \end{center}
\end{figure}

\section{Architectural overview of 'Harden' app}

The technology stack and architecture used for Harden is outlined in Figure \ref{fig:harden_architecture}, and described in detail below.

Harden uses the \pkg{shinyMobile} package for a reactive front-end, which is a layer on top of the \pkg{shiny} package and the \textit{framework7} HTML template \citep{framework72021}. The \pkg{shiny} package is a web application framework for R, created and maintained by RStudio, to reduce the complexity of creating interactive web applications without the need for HTML, CSS, or JavaScript knowledge \citep{cheng2021shiny}. The architecture of the \pkg{shiny} package is beyond the scope of this article, but is explained in depth elsewhere \citep{cheng2021shiny, kasprzak2020six}, along with the concept of reactivity \citep{grolemund_2015}.

\subsection{Database}

The \pkg{shiny} concept was extended to enable CRUD (Create, Read, Update, Delete) functionality, which was essential to allow participants to both input and read their variable data. All persistent user data was housed in a PostgreSQL database. The \pkg{DBI} and \pkg{RPostgres} packages were used to interface between the \pkg{shinyMobile} UI and the database \citep{wickham2021dbi, wickham2021rpostgres}.

To create and test the database locally, the command line tool \textit{psql} was used \citep{psql_2021}. A production database was initially set up on ElephantSQL, which at the time of writing provided free shared cloud hosting for a small PostgreSQL database up to 20 megabytes in size, with a limit of 5 concurrent connections for the free tier \citep{elephantsql2021}. Data was stored in separate tables based on its purpose, with the user ID set as the primary key for each table. For instance, one table housed all user settings, with each column containing an individual setting, while another table housed all variable data for each user. Using PostgreSQL for the backend provided the dual benefits of highly flexible database configurations coupled with rapid and scalable performance. 

For user authentication, the \pkg{polished} package was used, providing a customized sign-in page to the app, along with the ability to configure a database for maintaining user identities and logins \citep{tychobra2021polished}. The \pkg{polished} authentication layer was used to generate all user ID strings, which were stored as the primary key for each table in the PostgreSQL database.

\subsection{User interface and logic}

\begin{figure}[!ht]
    \begin{center}
        \includegraphics[width=1.0\textwidth]{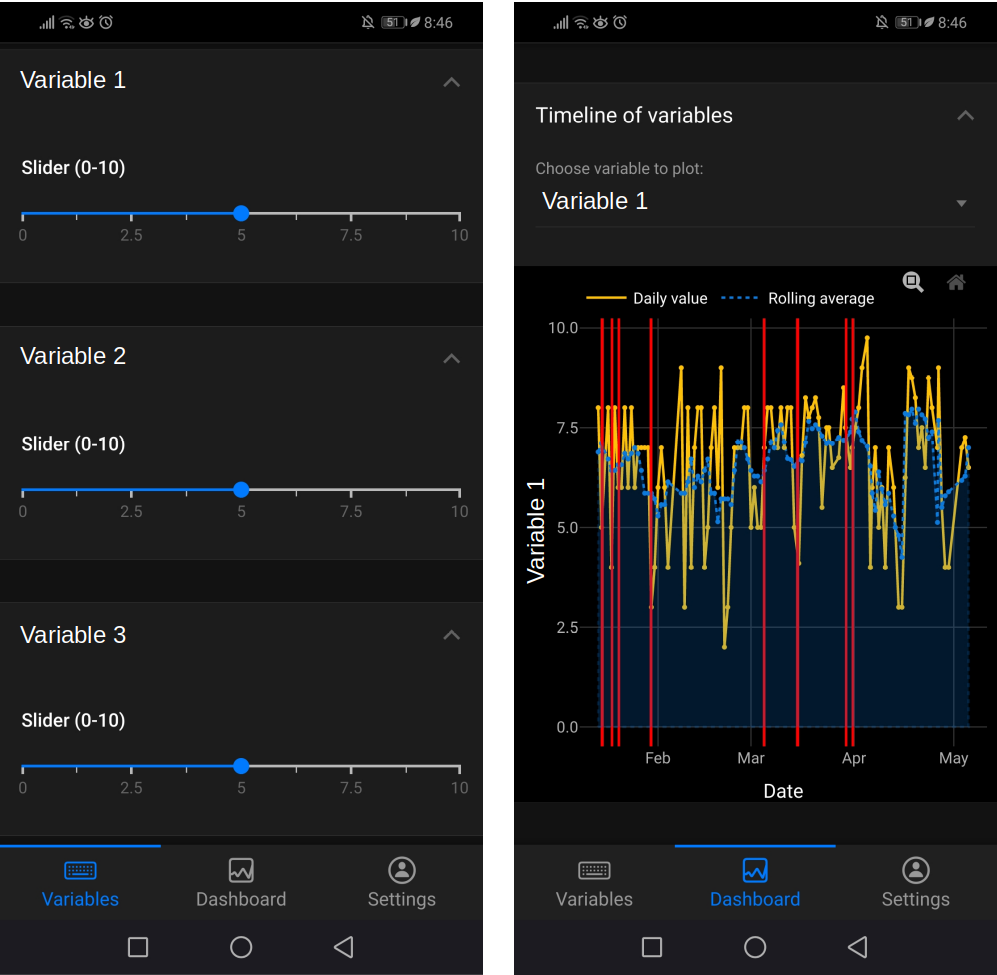}
        \caption{Example screenshots of user input tab (left) and graphs tab (right), taken from the Harden app running in Android as a PWA. All sliders and graphs are contained within accordion elements created with \code{shinyMobile::f7Accordion()}, and are updated reactively when the accordion item is opened, based on data queried from the PostgreSQL database. Graph elements are interactive, with features such as line selection and date filtering via two-finger zoom.}
        \label{fig:harden_images}
    \end{center}
\end{figure}

\pkg{shinyMobile} provides access to a multitude of \textit{framework7} widgets, several of which were utilised for Harden. For instance, \code{shinyMobile::f7Slider()} produces a slider bar for user input, and \code{shinyMobile::f7Accordion()} produces an 'accordion' widget, which allows different visual elements to be segmented into collapsible elements. The \code{shinyMobile::f7TabLayout()} template was used to divide the app into three tabs - one for user input, one for graphs, and one for user settings. All of these elements are demonstrated in Figure \ref{fig:harden_images}.

The \pkg{ggplot2} \citep{wickham2021ggplot2} and \pkg{plotly} \citep{sievert2021plotly} packages were combined to generate interactive plots, firstly by creating the \pkg{ggplot2} object, then passing it as an argument to \code{plotly::ggplotly()}. These plots were placed inside the accordion elements, and were made reactive, such that they loaded only when their accordion element was opened, thus reducing initial app load times. However, \pkg{plotly} does require loading of JavaScript libraries during creation of the first plot, which produces a slight delay for initial plot load times. Subsequent plots do not require the libraries to be re-loaded.

A dark theme is also available in \pkg{shinyMobile}. To match the dark theme, the \pkg{ggdark} package was used \citep{grantham2021ggdark}, which provides dark colour schemes for \pkg{ggplot2} graphs. 

The \pkg{shiny.pwa} package was used to convert the \pkg{shiny} app to a PWA, allowing it to run as a standalone fullscreen app with its own icon, on both mobile and desktop \citep{silva2021shiny}. Additionally, various packages (and their dependencies) were used to articulate the logic layer of the app, including:

\begin{itemize}
    \item The \pkg{dplyr}, \pkg{magrittr} and \pkg{tidyr} packages, for manipulating data in a tidy fashion \citep{wickham2021dplyr, bache2020magrittr, wickham2021tidyr}
    \item The \pkg{lubridate} and \pkg{zoo} packages, for manipulating date-time and time series data \citep{spinu2021lubridate, zeileis2021zoo}
    \item The \pkg{here} package, for generating relative filepaths and organising the code base \citep{muller2020here}
    \item The \pkg{sever} package, for creating a customized error message in cases where the connection to the server was lost \citep{coene2020sever}
\end{itemize}

\subsection{Accessory software}

Programming was mostly performed in the RStudio integrated development environment (IDE) \citep{rstudio2020ide}. All project files, including code files, images and related data, were regularly uploaded to a private GitHub repository for version control and backup purposes \citep{github2021}, using RStudio's inbuilt \textit{git} API \citep{chacon2014pro}. Prototypes of the app were hosted on the \textit{shinyapps.io} hosting platform \citep{shinyapps.io_2021}, also provided by RStudio and automated from within the RStudio IDE.

\subsection{Hosting}

In order to create a \pkg{shiny} app with full CRUD functionality, along with unique logins for each user, the server host must be able to provide an asynchronous platform for multiple users, to prevent cross-contamination of data updates between users. For this purpose, Google Cloud Run, which provides serverless hosting for containerized apps, was chosen as the host for Harden \citep{bisong2019overview}. Cloud Run provides a number of customizable options for its servers, including RAM and CPU limits. While Cloud Run can autoscale to zero server usage at times where there are no users, it also allows developers to specify a number of 'warm' instances running at all times. This prevents 'cold' starts if a user requests a container at a time when there are no spare containers running. For efficient autoscaling, it is recommended to have at least one container running warm on top of any other containers that are currently being used. This reduces start-up times for users who would otherwise deal with a cold start, but increases the runtime costs for Cloud Run \citep{google_2021}.

To deploy the application to Cloud Run, a Docker script was written to containerize the app and its requisite packages \citep{merkel2014docker}. The Docker script was then uploaded to the GitHub repository containing the code base for the app, with the Cloud Run instance being configured to trigger a new version build whenever a commit was pushed to the master GitHub repository branch. This allowed for easy version handling, and the option to fall back to an older version if critical bugs were found in the current version. It also ensured that users were kept up to date with the latest version, which was automatically propagated to all users at build time. Where packages were not strictly necessary to the functioning of Harden, they were excluded from the Dockerfile, in order to maintain a lightweight code base. Package versions were also specified in each Dockerfile, meaning that different app versions were not affected by unexpected package updates. Note that it is generally recommended for the developer to use a package such as \pkg{packrat} or \pkg{renv} to manage their project dependencies on their local environment \citep{ushey2021packrat, ushey2021renv}.

The database was initially hosted in a Google Compute Engine server located in Iowa in the United States, which helped to reduce latency between the Cloud Run server and the database. In a future version, the database will most likely be migrated to Google Cloud SQL (in partnership with Cloud Run), which will in theory provide further performance benefits due to lowered latency and network egress costs. 

\section{Performance}

R has a reputation for being less performant than other programming languages purpose-built for app creation \citep{lim2015r}. While this is debatable, the \pkg{shiny} package mostly bypasses the issue by running over a JavaScript layer, thus maximising reactive speed. In testing and production, the Harden app is acceptably responsive, with most boot times taking less than two seconds, and with near-instantaneous reactions to button presses. Simple graphs are calculated within a second, while more complicated algorithms can take up to a few seconds to process. Loading animations are included for graphs to provide users with information on loading times, by wrapping each \pkg{plotly} call inside a call to \code{shinycssloaders::withSpinner()}, from the \pkg{shinycssloaders} package \citep{attali2021css}.

Reactive UI elements, such as graphs and buttons, are calculated and rendered as HTML by the server before being passed to the phone's browser, which renders the HTML into visual elements. This process takes at most a few seconds for the most complex elements, particularly those that require SQL queries and further data processing prior to being rendered. The speed of calculation is dependent on the chosen server specifications, but even when using only 1 virtual CPU and the lowest amount of RAM necessary, most graphs were returned to the user in less than a second. 

One major performance bottleneck to take into account for rendering times is network latency, which is proportional to the user's distance from the nearest Cloud Run server. To reduce latency issues worldwide, a global load balancer is required, although this can significantly increase the cost of a Cloud Run subscription. While internet speeds and coverage continue to improve rapidly, a large proportion of the world's population is located at a significant distance from the closest Cloud Run data centre. This remains a disadvantage for the PWA method when compared to native mobile apps, which generally store data locally and hence aren't affected by network round times.  

In terms of features, it can be argued that any performance losses of the PWA method are neutralized by the advanced plotting capabilities provided by the \textit{CRAN} ecosystem. In cases where there are critical performance bottlenecks, programmers can use the \textit{Rcpp} package to incorporate C++ into their code \citep{eddelbuettel2011rcpp}.

\section{Discussion}

The strategy used to create the Harden app turned out to be successful, at least in terms of proof of concept. In terms of development time, the \pkg{shinyMobile} package is unique in that it offers a highly efficient methodology for generation of apps with sophisticated statistical and logical requirements, while also offering the potential for these apps to be put directly into production. If the use case for the app leans towards data science or statistical processing, then the \pkg{shinyMobile} framework may be superior to other programming languages for development purposes. This becomes especially true when one considers how crucial time to market has become in the software marketplace. If one can create a superior app as a solo developer with \pkg{shinyMobile} compared to a team of developers using Java and Swift, and at a faster rate, then this has the potential to foster creativity in a multitude of entrepreneurial fields - especially those adjacent to statistics.

While designing the app, it became apparent that the development process had to be approachable for a solo developer with limited programming experience. Remarkably, the technology stack detailed above allowed this app to be put into production in a mere seven months by a one-man development team. This feat can easily be bettered by more talented, hardworking and experienced programmers, but is indicative of the possibilities that the \pkg{shinyMobile} package has introduced. It would likely have taken years for the same solo developer to create a native app with identical functionality, that worked on both Android and iOS. 

The ability to work with R's graphing and data management libraries, particularly packages included in the \pkg{tidyverse} \citep{wickham2021tidyverse} such as \pkg{ggplot2} and \pkg{dplyr}, was a major boon to productivity, with many data manipulation tasks only requiring a single line of code in R. By comparison, to implement most of these tasks in a language such as Java or Swift would require substantially more code. However, this strategy can only be recommended for certain use cases. Prospective app developers are strongly encouraged to create a list of requirements for their app, and ensure that these requirements are met by the technology stack proposed in this paper, before replicating this strategy. The key limitations of this stack are listed below (although this list is not exhaustive). 

\section{Limitations}

\subsection{Internet connectivity}

The primary disadvantage of the \pkg{shinyMobile} technology is the requirement for users to be connected to a \pkg{shiny} server. While some PWAs have limited offline capabilities, a \pkg{shinyMobile} app relies heavily on the server to produce live data for the reactive UI, resulting in an immediate crash if the server is disconnected. As internet coverage improves worldwide, this problem is reduced over time, but app designers will need to assess whether there will be strong demand for the app in places of limited internet connectivity. 

\subsection{Native functionality}

A PWA, being a rendered web page, is unable to access most hardware functions on mobile, for security reasons. This means that certain native functions, such as scheduled notifications, are impossible to access through web applications. For the purposes of an EMA app, this is a significant disadvantage, as users need to be reminded to input their data on a daily basis at least \citep{shiffman2008ecological}. Two potential solutions include scheduled emails - potentially via the \pkg{blastula} package \citep{iannone2020blastula} - or housing the web app inside a WebView component in either a native Android or iOS app, and then scheduling notifications from the native app. However, these are convoluted workarounds, and a more straightforward solution would be welcomed. Depending on the importance of notifications to the app being created, it may be that developers will prefer development on native platforms to bypass this issue.

\subsection{Security}

The \pkg{shiny} and \pkg{shinyMobile} packages provide a streamlined way to reduce the complexity of web development into one language. This is an enormous advantage for statistical programmers with a limited understanding of internet technologies. However, it comes at the cost of introducing data security issues, as novice app developers explore untested methods of transferring data online \citep{kasprzak2020six}. 

In web security, it is vital to be aware of the concept of 'unknown unknowns'. Novice programmers are likely to make critical security errors, being ignorant of the dangers of poorly written code. In particular, statistical programmers who only have experience developing software on their local development environment are likely to be unfamiliar with modern web security needs. In other words, they are unable to fix vulnerabilities that they are not aware of \citep{kasprzak2020six}. As such, it is vitally important that all \pkg{shiny} developers who plan to transfer data across networks are trained in good practices, or that they recruit a data security expert to review their project.

Since RStudio's own \pkg{shiny} hosting service, \textit{shinyapps.io}, was not used for distributing Harden, additional security measures needed to be implemented to secure the Harden app from intruders over the network. These measures included the use of Secure Sockets Layer (SSL) certificates for encryption of packets over networks using the Hypertext Transfer Protocol Secure (HTTPS) communication protocol \citep{el2012most}, along with the sanitization of SQL queries for prevention of SQL injection attacks. The RStudio website provides guidance on how to create parameterized queries with the \code{DBI::dbSendQuery()} and \code{DBI::dbBind()} functions \citep{rstudio2021}. While a full discussion of security issues is beyond the scope of this article, it is vital that one is aware of the risks when designing their app.

\subsection{Cost}

For an app to have global reach, one should expect to pay a regular subscription fee to host their app on a web service such as Google Cloud Run or Amazon Web Services. This fee scales up per user. Pricing details for Google Cloud Run are provided on their website \citep{googlecloudrun}, and include details of a free trial, which will be beneficial for initial development and testing. Pricing will be a vital consideration for app developers, who may need to consider funding for research projects, or charging a subscription fee to users. For the latter option, the \pkg{polishedpayments} package \citep{merlino2021polishedpayments} can be used to create a paywall, in combination with the \pkg{polished} package. However, there is also a charge for using this service. It may be necessary for the developer to consider alternative cost-saving options, including creating their own authentication database or hosting their \pkg{shiny} server locally, if they cannot make their business model profitable while using \pkg{polished} in combination with Cloud Run. It is hoped that alternative solutions with lower margin costs will be made available in the near future, as more developers explore this space.

% Alternatives for authentication include shinymanager, shinyauthr, shinyproxy, shiny-auth0, ......

\section{Summary}

The \pkg{shinyMobile} package provides a unique entry point for statistical programmers into the app-development market. Programmers with some experience in R are likely to find that creating a \pkg{shinyMobile} app is far more time efficient than creating the equivalent app in a native mobile language. The technology stack outlined in this paper is just one example of how a CRUD-capable PWA can be put into production, coded mostly in R. Necessary future developments in this space include finding workarounds for reduced native hardware support, as well as improving documentation and support, such that it becomes easier to link cloud providers and databases with the main R code base in a more reproducible fashion. Overall, this is a rapidly evolving field with vast potential, and the \pkg{shinyMobile} package may prove to be a significant force of innovation in the field of statistical mobile app development.

%% file: RJwrapper.bbl
\begin{thebibliography}{45}
\providecommand{\natexlab}[1]{#1}
\providecommand{\url}[1]{\texttt{#1}}
\expandafter\ifx\csname urlstyle\endcsname\relax
  \providecommand{\doi}[1]{doi: #1}\else
  \providecommand{\doi}{doi: \begingroup \urlstyle{rm}\Url}\fi

\bibitem[Kasprzak et~al.(2020)Kasprzak, Mitchell, Kravchuk, and
  Timmins]{kasprzak2020six}
Peter Kasprzak, Lachlan Mitchell, Olena Kravchuk, and Andy Timmins.
\newblock Six years of shiny in research-collaborative development of web tools
  in r.
\newblock \emph{R JOURNAL}, 12\penalty0 (2):\penalty0 20--42, 2020.

\bibitem[Corral et~al.(2012)Corral, Janes, and Remencius]{corral2012potential}
Luis Corral, Andrea Janes, and Tadas Remencius.
\newblock Potential advantages and disadvantages of multiplatform development
  frameworks--a vision on mobile environments.
\newblock \emph{Procedia Computer Science}, 10:\penalty0 1202--1207, 2012.

\bibitem[Chambers(2000)]{chambers2000users}
John~M Chambers.
\newblock Users, programmers, and statistical software.
\newblock \emph{Journal of Computational and Graphical Statistics}, 9\penalty0
  (3):\penalty0 404--422, 2000.

\bibitem[Heitk{\"o}tter et~al.(2013)Heitk{\"o}tter, Majchrzak, and
  Kuchen]{heitkotter2013cross}
Henning Heitk{\"o}tter, Tim~A Majchrzak, and Herbert Kuchen.
\newblock Cross-platform model-driven development of mobile applications with
  md2.
\newblock In \emph{Proceedings of the 28th Annual ACM Symposium on Applied
  Computing}, pages 526--533, 2013.

\bibitem[Bi{\o}rn-Hansen et~al.(2017)Bi{\o}rn-Hansen, Majchrzak, and
  Gr{\o}nli]{biorn2017progressive}
Andreas Bi{\o}rn-Hansen, Tim~A Majchrzak, and Tor-Morten Gr{\o}nli.
\newblock Progressive web apps: The possible web-native unifier for mobile
  development.
\newblock In \emph{International Conference on Web Information Systems and
  Technologies}, volume~2, pages 344--351. SciTePress, 2017.

\bibitem[Granjon et~al.(2021)Granjon, Perrier, and
  Rudolf]{granjon2021shinymobile}
David Granjon, Victor Perrier, and Isabelle Rudolf.
\newblock \emph{shinyMobile: Mobile Ready 'shiny' Apps with Standalone
  Capabilities}, 2021.
\newblock https://github.com/RinteRface/shinyMobile,
  https://rinterface.github.io/shinyMobile/.

\bibitem[Elliott and Elliott(2020)]{elliott2020developing}
Matthew~S Elliott and Lisa~M Elliott.
\newblock Developing r shiny web applications for extension education.
\newblock \emph{Applied Economics Teaching Resources (AETR)}, 2\penalty0
  (4):\penalty0 TBD--TBD, 2020.

\bibitem[Chang et~al.(2021)Chang, Cheng, Allaire, Sievert, Schloerke, Xie,
  Allen, McPherson, Dipert, and Borges]{cheng2021shiny}
Winston Chang, Joe Cheng, JJ~Allaire, Carson Sievert, Barret Schloerke, Yihui
  Xie, Jeff Allen, Jonathan McPherson, Alan Dipert, and Barbara Borges.
\newblock \emph{shiny: Web Application Framework for R}, 2021.
\newblock URL \url{https://CRAN.R-project.org/package=shiny}.
\newblock R package version 1.6.0.

\bibitem[Kharlampidi(2021)]{framework72021}
Vladimir Kharlampidi.
\newblock Full featured framework for building ios, android \&amp; desktop
  apps, 2021.
\newblock URL \url{https://framework7.io/}.

\bibitem[Grolemund(2015)]{grolemund_2015}
Garrett Grolemund, May 2015.
\newblock URL
  \url{https://shiny.rstudio.com/articles/understanding-reactivity.html}.

\bibitem[{R Special Interest Group on Databases (R-SIG-DB)} et~al.(2021){R
  Special Interest Group on Databases (R-SIG-DB)}, Wickham, and
  Müller]{wickham2021dbi}
{R Special Interest Group on Databases (R-SIG-DB)}, Hadley Wickham, and Kirill
  Müller.
\newblock \emph{DBI: R Database Interface}, 2021.
\newblock URL \url{https://CRAN.R-project.org/package=DBI}.
\newblock R package version 1.1.1.

\bibitem[Wickham et~al.(2021{\natexlab{a}})Wickham, Ooms, and
  Müller]{wickham2021rpostgres}
Hadley Wickham, Jeroen Ooms, and Kirill Müller.
\newblock \emph{RPostgres: 'Rcpp' Interface to 'PostgreSQL'},
  2021{\natexlab{a}}.
\newblock URL \url{https://CRAN.R-project.org/package=RPostgres}.
\newblock R package version 1.3.2.

\bibitem[psq(2021)]{psql_2021}
Psql, Aug 2021.
\newblock URL \url{https://www.postgresql.org/docs/13/app-psql.html}.

\bibitem[Hörberg(2021)]{elephantsql2021}
Carl Hörberg.
\newblock Elephantsql - postgresql as a service, 2021.
\newblock URL \url{https://www.elephantsql.com/}.

\bibitem[Merlino et~al.(2021)Merlino, Howard, and Briggs]{tychobra2021polished}
Andy Merlino, Patrick Howard, and Jimmy Briggs.
\newblock \emph{polished: Authentication, User Administration, and Hosting for
  'shiny' Apps}, 2021.
\newblock https://github.com/tychobra/polished, https://polished.tech.

\bibitem[Wickham et~al.(2021{\natexlab{b}})Wickham, Chang, Henry, Pedersen,
  Takahashi, Wilke, Woo, Yutani, and Dunnington]{wickham2021ggplot2}
Hadley Wickham, Winston Chang, Lionel Henry, Thomas~Lin Pedersen, Kohske
  Takahashi, Claus Wilke, Kara Woo, Hiroaki Yutani, and Dewey Dunnington.
\newblock \emph{ggplot2: Create Elegant Data Visualisations Using the Grammar
  of Graphics}, 2021{\natexlab{b}}.
\newblock URL \url{https://CRAN.R-project.org/package=ggplot2}.
\newblock R package version 3.3.5.

\bibitem[Sievert et~al.(2021)Sievert, Parmer, Hocking, Chamberlain, Ram,
  Corvellec, and Despouy]{sievert2021plotly}
Carson Sievert, Chris Parmer, Toby Hocking, Scott Chamberlain, Karthik Ram,
  Marianne Corvellec, and Pedro Despouy.
\newblock \emph{plotly: Create Interactive Web Graphics via 'plotly.js'}, 2021.
\newblock URL \url{https://CRAN.R-project.org/package=plotly}.
\newblock R package version 4.9.3.

\bibitem[Grantham(2019)]{grantham2021ggdark}
Neal Grantham.
\newblock \emph{ggdark: Dark Mode for 'ggplot2' Themes}, 2019.
\newblock URL \url{https://CRAN.R-project.org/package=ggdark}.
\newblock R package version 0.2.1.

\bibitem[Silva(2021)]{silva2021shiny}
Pedro Silva.
\newblock \emph{shiny.pwa: Progressive Web App Support for Shiny}, 2021.
\newblock URL \url{https://github.com/pedrocoutinhosilva/shiny.pwa}.
\newblock R package version 0.2.0.

\bibitem[Wickham et~al.(2021{\natexlab{c}})Wickham, François, Henry, and
  Müller]{wickham2021dplyr}
Hadley Wickham, Romain François, Lionel Henry, and Kirill Müller.
\newblock \emph{dplyr: A Grammar of Data Manipulation}, 2021{\natexlab{c}}.
\newblock URL \url{https://CRAN.R-project.org/package=dplyr}.
\newblock R package version 1.0.5.

\bibitem[Bache and Wickham(2020)]{bache2020magrittr}
Stefan~Milton Bache and Hadley Wickham.
\newblock \emph{magrittr: A Forward-Pipe Operator for R}, 2020.
\newblock URL \url{https://CRAN.R-project.org/package=magrittr}.
\newblock R package version 2.0.1.

\bibitem[Wickham(2021)]{wickham2021tidyr}
Hadley Wickham.
\newblock \emph{tidyr: Tidy Messy Data}, 2021.
\newblock URL \url{https://CRAN.R-project.org/package=tidyr}.
\newblock R package version 1.1.3.

\bibitem[Spinu et~al.(2021)Spinu, Grolemund, and Wickham]{spinu2021lubridate}
Vitalie Spinu, Garrett Grolemund, and Hadley Wickham.
\newblock \emph{lubridate: Make Dealing with Dates a Little Easier}, 2021.
\newblock URL \url{https://CRAN.R-project.org/package=lubridate}.
\newblock R package version 1.7.10.

\bibitem[Zeileis et~al.(2021)Zeileis, Grothendieck, and Ryan]{zeileis2021zoo}
Achim Zeileis, Gabor Grothendieck, and Jeffrey~A. Ryan.
\newblock \emph{zoo: S3 Infrastructure for Regular and Irregular Time Series
  (Z's Ordered Observations)}, 2021.
\newblock URL \url{https://CRAN.R-project.org/package=zoo}.
\newblock R package version 1.8-9.

\bibitem[Müller(2020)]{muller2020here}
Kirill Müller.
\newblock \emph{here: A Simpler Way to Find Your Files}, 2020.
\newblock URL \url{https://CRAN.R-project.org/package=here}.
\newblock R package version 1.0.1.

\bibitem[Coene(2020)]{coene2020sever}
John Coene.
\newblock \emph{sever: Customise 'Shiny' Disconnected Screens and Error
  Messages}, 2020.
\newblock URL \url{https://CRAN.R-project.org/package=sever}.
\newblock R package version 0.0.6.

\bibitem[{RStudio Team}(2020)]{rstudio2020ide}
{RStudio Team}.
\newblock \emph{RStudio: Integrated Development Environment for R}.
\newblock RStudio, PBC., Boston, MA, 2020.
\newblock URL \url{http://www.rstudio.com/}.

\bibitem[Github(2021)]{github2021}
Github.
\newblock Where the world builds software, 2021.
\newblock URL \url{https://github.com/}.

\bibitem[Chacon and Straub(2014)]{chacon2014pro}
Scott Chacon and Ben Straub.
\newblock \emph{Pro git}.
\newblock Apress, 2014.

\bibitem[RStudio(2021{\natexlab{a}})]{shinyapps.io_2021}
RStudio.
\newblock Shinyapps.io, 2021{\natexlab{a}}.
\newblock URL \url{https://www.shinyapps.io/}.

\bibitem[Bisong(2019)]{bisong2019overview}
Ekaba Bisong.
\newblock An overview of google cloud platform services.
\newblock \emph{Building Machine Learning and Deep Learning Models on Google
  Cloud Platform}, pages 7--10, 2019.

\bibitem[Google(2021{\natexlab{a}})]{google_2021}
Google.
\newblock 3 ways to optimize cloud run response times | google cloud blog,
  2021{\natexlab{a}}.
\newblock URL
  \url{https://cloud.google.com/blog/topics/developers-practitioners/3-ways-optimize-cloud-run-response-times}.

\bibitem[Merkel(2014)]{merkel2014docker}
Dirk Merkel.
\newblock Docker: lightweight linux containers for consistent development and
  deployment.
\newblock \emph{Linux journal}, 2014\penalty0 (239):\penalty0 2, 2014.

\bibitem[Ushey et~al.(2021)Ushey, McPherson, Cheng, Atkins, and
  Allaire]{ushey2021packrat}
Kevin Ushey, Jonathan McPherson, Joe Cheng, Aron Atkins, and JJ~Allaire.
\newblock \emph{packrat: A Dependency Management System for Projects and their
  R Package Dependencies}, 2021.
\newblock URL \url{https://CRAN.R-project.org/package=packrat}.
\newblock R package version 0.6.0.

\bibitem[Ushey(2021)]{ushey2021renv}
Kevin Ushey.
\newblock \emph{renv: Project Environments}, 2021.
\newblock URL \url{https://CRAN.R-project.org/package=renv}.
\newblock R package version 0.14.0.

\bibitem[Lim and Tjhi(2015)]{lim2015r}
Aloysius Lim and William Tjhi.
\newblock \emph{R High Performance Programming}.
\newblock Packt Publishing Ltd, 2015.

\bibitem[Sali and Attali(2020)]{attali2021css}
Andras Sali and Dean Attali.
\newblock \emph{shinycssloaders: Add Loading Animations to a 'shiny' Output
  While It's Recalculating}, 2020.
\newblock URL \url{https://CRAN.R-project.org/package=shinycssloaders}.
\newblock R package version 1.0.0.

\bibitem[Eddelbuettel and Fran\c{c}ois(2011)]{eddelbuettel2011rcpp}
Dirk Eddelbuettel and Romain Fran\c{c}ois.
\newblock {Rcpp}: Seamless {R} and {C++} integration.
\newblock \emph{Journal of Statistical Software}, 40\penalty0 (8):\penalty0
  1--18, 2011.
\newblock \doi{10.18637/jss.v040.i08}.
\newblock URL \url{https://www.jstatsoft.org/v40/i08/}.

\bibitem[Wickham(2019)]{wickham2021tidyverse}
Hadley Wickham.
\newblock \emph{tidyverse: Easily Install and Load the 'Tidyverse'}, 2019.
\newblock URL \url{https://CRAN.R-project.org/package=tidyverse}.
\newblock R package version 1.3.0.

\bibitem[Shiffman et~al.(2008)Shiffman, Stone, and
  Hufford]{shiffman2008ecological}
Saul Shiffman, Arthur~A Stone, and Michael~R Hufford.
\newblock Ecological momentary assessment.
\newblock \emph{Annu. Rev. Clin. Psychol.}, 4:\penalty0 1--32, 2008.

\bibitem[Iannone and Cheng(2020)]{iannone2020blastula}
Richard Iannone and Joe Cheng.
\newblock \emph{blastula: Easily Send HTML Email Messages}, 2020.
\newblock URL \url{https://CRAN.R-project.org/package=blastula}.
\newblock R package version 0.3.2.

\bibitem[El-Hajj(2012)]{el2012most}
Wassim El-Hajj.
\newblock The most recent ssl security attacks: origins, implementation,
  evaluation, and suggested countermeasures.
\newblock \emph{Security and Communication Networks}, 5\penalty0 (1):\penalty0
  113--124, 2012.

\bibitem[RStudio(2021{\natexlab{b}})]{rstudio2021}
RStudio, 2021{\natexlab{b}}.
\newblock URL \url{https://db.rstudio.com/best-practices/run-queries-safely/}.

\bibitem[Google(2021{\natexlab{b}})]{googlecloudrun}
Google.
\newblock Cloud run pricing, 2021{\natexlab{b}}.
\newblock URL \url{https://cloud.google.com/run/pricing}.

\bibitem[Merlino(2021)]{merlino2021polishedpayments}
Andy Merlino.
\newblock \emph{polishedpayments: Easily add a Stripe subscription to your
  Shiny app}, 2021.
\newblock https://github.com/tychobra/polishedpayments, https://polished.tech.

\end{thebibliography}
